\documentclass[aps,superscriptaddress,amsfonts,amssymb,amsmath,showpacs,nofootinbib, twocolumn, floatfix]{revtex4-1}
\usepackage{graphicx}
\usepackage[squaren]{SIunits}

\newcommand{\Eref}[1]{Eq.~\eqref{#1}} 
\newcommand{\Eqnsref}[2]{Equations~\eqref{#1} and~\eqref{#2}} 
\newcommand{\Fref}[1]{Fig.~\ref{#1}} 
\newcommand{\Sref}[1]{Sec.~\ref{#1}} 
\newcommand{\Tref}[1]{Table~\ref{#1}} 

\begin{document}

\title{Fundamental oscillation modes of neutron stars: validity of universal
relations}

\author{Cecilia Chirenti} 
\email{e-mail: cecilia.chirenti@ufabc.edu.br} 
\affiliation{Centro de Matem\'atica, Computa\c c\~ao e Cogni\c c\~ao, UFABC, 09210-170 Santo Andr\'e-SP, Brazil}

\author{Gibran H. de Souza} 
\affiliation{Instituto de F\'isica Gleb Wataghin, UNICAMP, 13083-859, Campinas-SP, Brazil}

\author{Wolfgang Kastaun} 
\affiliation{Physics Department, University of Trento, 
via Sommarive 14, I-38123 Trento, Italy}

\begin{abstract}
We study the $f$-mode frequencies and damping times of nonrotating neutron stars (NS) in general relativity (GR)  by solving the linearized perturbation equations, with the aim to establish ``universal'' relations that depend only weakly on the equations of state (EOS). Using a more comprehensive set of EOSs, we re-examine some proposed 
empirical relations 
that describe the $f$-mode parameters in terms of mass and radius of the neutron star (NS), and we test a more recent proposal for expressing the $f$-mode parameters as quadratic functions of the effective compactness. Our extensive results for each equation of state considered allow us to study the accuracy of each proposal. 
In particular, the empirical relation proposed in the literature
for the damping time in terms of the mass and radius deviates considerably
from our results.
We introduce a new universal relation for the product of the $f$-mode frequency and damping time as a function of the (ordinary) compactness, which proved to be more accurate.
The more recently proposed relations using the effective compactness, on the other hand, also fit our data accurately.
Our results show that the maximum oscillation frequency depends strongly on the EOS,
such that the measurement of a high oscillation frequency would rule out several
EOSs. 
Lastly, we compare the exact mode frequencies to those obtained in the Cowling 
approximation, and also to results obtained with a nonlinear evolution code, 
validating the implementations of the different approaches.
\end{abstract}


\maketitle

\section{Introduction}
\label{sec:intro}

\begin{figure}[!htb]
\begin{center}
\includegraphics[width=0.98\columnwidth]{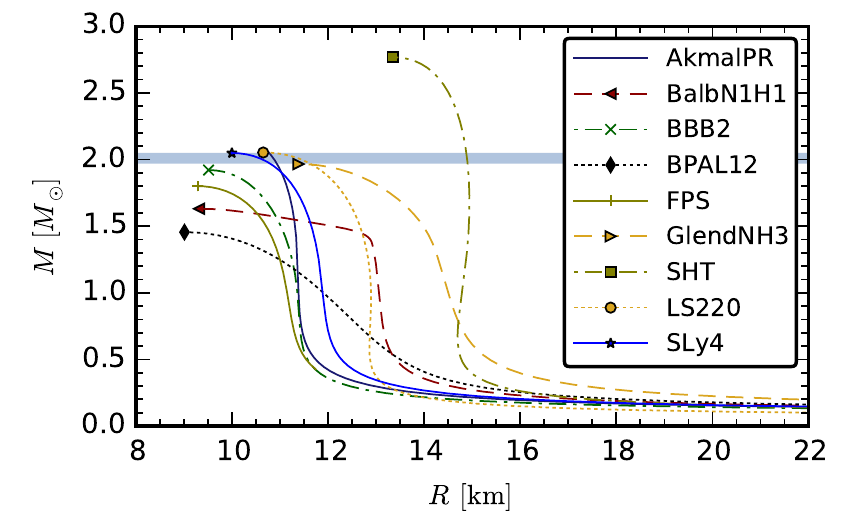}
\end{center}
\caption{Gravitational mass $M$ versus circumferential radius $R$ for the 
set of EOS used in this work. The shaded area shows the mass range 
measured by \cite{Antoniadis13} for the pulsar PSR J0348+0432.
The symbols mark the maximum mass models.}
\label{fig:M-R}
\end{figure}

Neutron stars are among the most interesting celestial objects 
since their description requires both general relativity and 
nuclear physics. 
The state of matter in the core of NSs is not accessible to any 
terrestrial experiments, and thus provides a unique laboratory to 
test theoretical predictions for matter at high densities and 
relatively low temperatures (compared to particle collisions 
producing comparable energy densities).
In particular, there exist many different models predicting the 
EOS of NS matter, see \cite{Cole, SHT, LS220, SLy1, 
SLy2, GlendNH3, FPS, BPAL12, BBB2, BablbN1H1, AkmalPR}, among others. 
A possibility to constrain those models from NS observations is 
given by the fact that general relativity predicts a maximum mass 
for non-rotating and uniformly rotating NSs, which depends on the 
EOS. 
Recently, \cite{Demorest} and \cite{Antoniadis13} discovered
NSs with masses as high as $1.97 \, M_{\odot}$ and $2.01 \, M_{\odot}$,
respectively. 
This already ruled out several EOSs, as shown in \Fref{fig:M-R}.
A simultaneous observation of mass and radius of a slowly rotating NS 
would serve the same purpose.

Given a set of observed quantities, constraining the EOS obviously 
requires relations which do depend strongly on the EOS. 
On the other hand, relations which do \emph{not} depend on the EOS 
(or only weakly), are highly useful as well, since they would allow
to constrain further unknown parameters without knowledge of the EOS.
Such universal relations have already been established in the context
of binary NS evolution between moment of inertia, the tidal 
deformability (Love number) and the quadrupole moment, both for 
regular NSs and quark stars, see \cite{Yagi1,Yagi2,Maselli}.
Weakly EOS dependent relations between compactness and normalized 
moment of inertia have been found by \cite{Bejger,Lattimerschutz05}.

Other studies \cite{Andersson, Benhar, Lau, Tsui} have investigated 
various empirical universal expressions for oscillation frequencies 
and damping timescales.
Among the various types of oscillation modes, the main focus has 
been on modes for which pressure is the restoring force, in 
particular the fundamental modes ($f$ modes). 
Knowledge of the $f$ modes of nonrotating stars also provides 
estimates for the $f$-mode properties of slowly rotating stars, for 
the case of uniform rotation \cite{Yoshida} and also for differential
rotation \cite{Chirenti}.

Some years ago, Andersson and Kokkotas \cite{Andersson} proposed an
empirical relation for the $f$-mode oscillation frequency $\omega$, 
based on Newtonian theory of stellar perturbations. 
They observed that in full GR, $\omega$ depends almost linearly on 
the square root of the average density. 
Another relation, based on estimates using the quadrupole formula, 
was established for the damping time due to gravitational wave 
emission, $\tau$.
Later, Benhar {\it et al.} \cite{Benhar} presented further results 
that included more and newer equations of state, updating the fits 
from \cite{Andersson}.
The average frequencies $\omega$ were systematically lower than the 
one for the old EOS sample, which they attributed to the fact that 
the new sample included stiffer EOSs. 

This highlights an important point: all the universal relations are 
not truly EOS-independent, but restrain parameters in a relatively 
narrow band which depends on the EOSs taken into account.
Using this band to interpret observational data implicitly  
assumes that the true EOS is similar to one of the EOSs considered. 
It is therefore important to include a large range of EOSs in the 
sample, to prevent underestimating the residual uncertainties due to 
the EOS.
For this reason, we do not exclude EOSs already ruled out by the 
observation \cite{Antoniadis13}.
 
More recently, Lau {\it et al.} \cite{Lau} improved on the results 
from \cite{Tsui} by using quadratic fits for both $\omega$ and 
$\tau$. Further, they use the ``effective compactness'' 
$\eta \equiv \sqrt{M^3/I}$ as independent variable, where $I$ is the 
moment of inertia. In contrast to measures involving the surface 
radius, $\eta$ is determined by the bulk properties of the NS. 
Note the moment of inertia could be constrained for isolated stars by 
observations of spindown and energy output due to magnetic braking, 
see \cite{Bejger}. It might also be measured from high precision observations
of spin-orbit couplings in a binary pulsar system, see \cite{Lattimerschutz05}.

Observing oscillations of isolated NSs might in principle be possible 
through conventional astronomy if they are equipped with a magnetic 
field. 
The direct production of electromagnetic emissions due to NS (and BH) 
perturbations was investigated in \cite{Sotani2013, Sotani2014}. 
Stellar oscillations might also modulate EM emissions produced by 
other processes. 
For example, possible signatures of NS oscillations have already been 
observed in the luminosity curves of magnetar giant flares 
\cite{Watts2006}. 
However, the low frequencies rule out fundamental modes as the cause. 
Instead, the modulation was attributed to oscillations of the 
magnetar crust and magnetic field, see \cite{Colaiuda2009} and the 
references therein.

Another observational channel is given by the gravitational waves 
emitted by oscillating NSs, which have been the subject of many 
studies over the years (see \cite{Kokkotas,Nollert,Allen} and 
references therein).
The detection range is however limited by the fact that oscillations 
above a certain amplitude are strongly damped by nonlinear effects 
\cite{Kastaun10}.
Moreover, possible excitation mechanisms most likely already saturate 
at much lower amplitudes. 
The most famous mechanism is the CFS instability \cite{CFS}, which is
restricted to rapidly rotating NSs (for the case of $f$ modes), and 
probably suppressed by superfluid or viscous effects outside the 
instability window described in \cite{Doneva13}. 
More speculative excitation mechanisms are phase transitions of the 
EOS \cite{Abdikamalov09}, or resonant excitation in eccentric 
binaries, proposed by \cite{Pons} (although it seems questionable 
whether the binary stays in the resonant window sufficiently long).

The most promising source of detectable GW from NSs is given by the 
hyper- or supramassive neutron stars which can be formed in binary NS 
mergers \cite{Baiotti}. 
However, those are rapidly and differentially rotating. 
Further, they are hot and not in $\beta$-equilibrium, and thus do not 
follow a simple barotropic EOS. 
Due to these additional degrees of freedom, universal relations found 
for slowly rotating cold stars are not directly applicable.
Still, a better understanding of the simple nonrotating case is 
certainly beneficial for the development of more sophisticated models 
needed to describe hypermassive NSs.

Our first aim in this paper is to provide accurate results for the 
$f$ modes in a wide range of masses and the EOSs given in  
\Tref{tableEOS}, which we hope will be useful for the community. 
In spite of being widely used in this field, properties of the 
$f$ modes are not available in the literature for some of the EOSs.

\begin{table}[h]
\caption{The equations of state used in this paper. All of the EOS, except for the 
last two, are distributed with the publicly available Lorene code \cite{Lorene}.
The LS220 and SHT EOSs include temperature and composition dependency; we use
the barotropic EOSs obtained by imposing zero-temperature and $\beta$-equilibrium.
}
\label{tableEOS}
 \begin{ruledtabular}
  \begin{tabular}{ll||ll||ll}
     EOS & Ref. & EOS & Ref. & EOS & Ref.\\
     \hline
     AkmalPR  & \cite{AkmalPR}   & BPAL12   & \cite{BPAL12}   & SLy4  & \cite{SLy1,SLy2} \\
     BalbN1H1 & \cite{BablbN1H1} & FPS      & \cite{FPS}      & LS220 & \cite{LS220}     \\
     BBB2     & \cite{BBB2}      & GlendNH3 & \cite{GlendNH3} & SHT   & \cite{SHT}       \\
 \end{tabular}
 \end{ruledtabular}
\end{table}

Our second aim is to update the empirical relations studied in \cite{Andersson} and \cite{Benhar}, and test the proposed universal relations from \cite{Lau} for the newer EOS we consider here. This analysis is presented in \Sref{sec:results}.

In \Sref{sec:compare}, we compare our linear results
to nonlinear, fully relativistic 3D simulations of selected models. We also study
the (in)accuracy of the relativistic Cowling approximation for selected EOSs.
Finally, we present our concluding remarks in \Sref{sec:conclusions}.
Numerical results for all equations of state considered here are provided 
in the Appendix.

\section{Numerical results for the $f$-mode frequencies and damping times}
\label{sec:results}

Our numerical scheme for obtaining the $f$-mode frequencies and damping times solves the set of perturbation equations derived in the original papers by Lindblom \& Detweiler \cite{Lindblom1983,Lindblom1985}, based on the pioneering work by Thorne and collaborators \cite{Thorne1967}-\cite{Thorne1973}. Our algorithm follows closely the outline given in \cite{Lindblom1983,Lindblom1985}, and some preliminary results obtained for a polytropic EOS were presented in \cite{Patrick}. 
We solve the eigenvalue problem using the Newtonian $f$-mode frequency as
an initial guess. The radial perturbation equations inside the star,
written as a system of 4 coupled ODEs, are solved with a standard 4th order
Runge-Kutta method. The solution obtained is then matched to the numerical
solution of the Zerilli equation at the surface and to the asymptotic
solution of the Zerilli equation at a large enough radius. Finally, we use
a shooting method to refine the initial guess and determine the complex
eigenfrequency, by imposing the purely outgoing wave condition at infinity.

Our code assumes that the EOS of the background model is also valid to describe
dynamic perturbations. This is an approximation since the system can deviate
from $\beta$-equilibrium. To estimate the importance of this error, we compare
the speed of sound for two corner cases:
$c_s^\beta$ is the soundspeed assuming that $\beta$-equilibrium is
always satisfied, and $c_s^Y$ is the soundspeed assuming that
the electron fraction is fixed along fluid worldlines.  
We define an average relative error of the soundspeed by 
$\bar{\delta} c_s = \sqrt{(\int \delta c_s^2 \rho \,\mathrm{d}\rho)
/ (\int \rho \,\mathrm{d}\rho)}$, with $\delta c_s = 1-c_s^\beta/c_s^Y$.
Of our EOS sample only 
the LS220 and SHT EOSs provide values out of $\beta$-equilibrium.
For both, we find $\bar{\delta} c_s \approx 1\%$. 
Since stellar oscillations are strongly related to the sound crossing
time, and assuming the frequencies are mainly determined by the bulk of
the star, we use $\bar{\delta} c_s$ also as a rough estimate for the
corresponding differences of the oscillation frequencies.
As an independent measure, we note
radial oscillation frequencies of NS have been compared in \cite{Galeazzi2013} 
for the two corner cases above, using the same EOSs. The resulting frequencies 
differ by ${\approx}1\%$, in agreement with the estimate above. 
Since the GW luminosity depends on the frequency, we 
can expect similar differences for the damping time.
For the rest of this article, we will ignore the influence of the 
$\beta$-equilibrium assumption.
The purely numerical error in the determination of the $f$-mode parameters 
is  approximately $0.1\%$ for $\omega$ and $2\%$ for $\tau$.
This was estimated by running convergence tests with 5000, 10000, 
and 20000 points for different models spanning the whole mass range.

\begin{figure*}[!htb]
\begin{center}
\includegraphics[width=\textwidth]{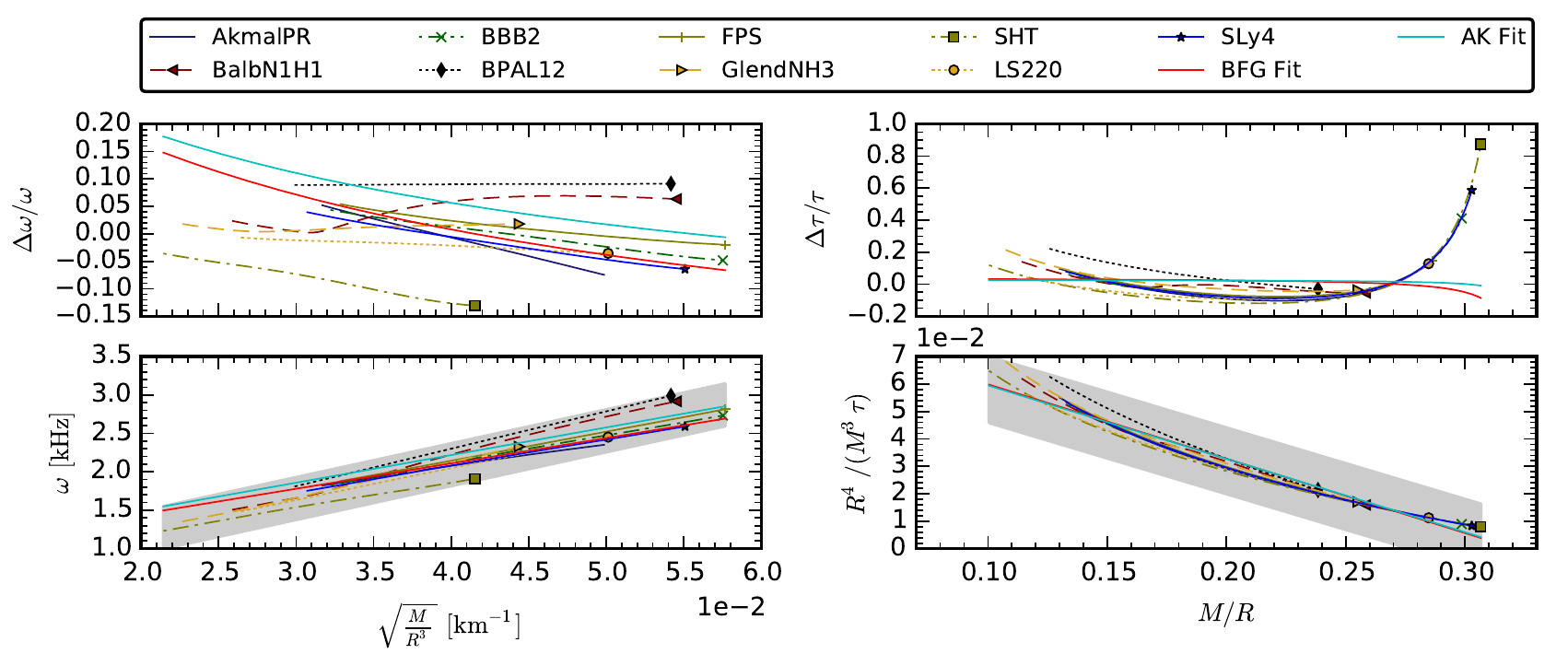}
\end{center}
\caption{
Properties of the $f$ mode in terms of mass and radius, and EOS-independent 
approximations.
The left panels show the relation \Eref{eq:fit1} for the frequency, 
the right panels the relation \Eref{eq:fit2} for the damping time.
The lower panels show the exact results for each EOS as well as  the fits
proposed in \cite{Benhar} (BFG) and \cite{Andersson} (AK). The shaded regions 
show the 
confidence bands (see main text) around our fits.
The upper panels show the remaining EOS dependency in terms of the 
residuals of our fit, as well as the differences to the AK and BFG fits.
}
\label{fig:fit1}
\end{figure*}

In the following, we present our numerical results and compare them with the empirical relations presented by Andersson \& Kokkotas \cite{Andersson}, Benhar {\it et al.} \cite{Benhar} and Lau {\it et al.} \cite{Lau}. Our selection of EOSs given in \Tref{tableEOS} 
differs from the ones used in those works in that we have more EOSs than \cite{Benhar} and newer EOSs that were not considered in \cite{Andersson} and \cite{Lau} (we do not consider here some of their older EOSs nor quark matter). We also include the LS220 and SHT EOSs which are frequently used in merger simulations.
For each EOS, we set up models in a large mass range, starting from $1 M_{\bigodot}$
up to the maximum mass. We sample the corresponding central energy density range 
linearly, using 101 data points for each EOS.
Note for the AkmalPR EOS, the central soundspeed becomes superluminal at a density
below the one of the maximum mass model. We excluded all the causality-violating
models.
Following \cite{Andersson} and \cite{Benhar}, we used the data from all the EOS to fit the empirical relations below, which relate the $f$-mode frequency $\omega$ to the square root of the average density $M/R^3$, and the damping time $\tau$ to the compactness $M/R$ of the neutron star:
\begin{align}
\omega &= a_1 + b_1 \sqrt{\frac{M}{R^3}} \,, 
\label{eq:fit1} \\
\frac{R^4}{M^3 \tau} &= 
a_2 + b_2\frac{M}{R} \,.
\label{eq:fit2}
\end{align}
These relations are shown in \Fref{fig:fit1}. As one can see, each EOS 
individually satisfies a linear relation given by \Eref{eq:fit1} to good accuracy.
There is however a considerable spread between the different EOSs.
We fitted \Eref{eq:fit1} to all models combined to get an average relation.
Since the spread is of systematic nature, the formal 
statistical errors of the fit are meaningless. We stress that the fit result itself
is also ambiguous, since it depends on the selection of models and there is no ``true'' 
value. Our aim is to establish a band around the fitted 
relation which contains all our results, and which can be used to constrain 
observational data without detailed knowledge of the EOS, only assuming that 
it is similar to one of the EOSs from our sample.
To define this ``confidence band'', we simply use the largest residual of the fit
as systematic error of the constant offset $a_1$.
Our fit results for \Eref{eq:fit1} are given in \Tref{tab:fit1_ind_eos}.
The table also contains the results of the same fit applied to each EOS individually,
which can be used to estimate oscillation frequencies for a given EOS to good
accuracy.
The fits from \cite{Andersson} and \cite{Benhar}, also shown \Fref{fig:fit1},
are also contained within the confidence band that envelopes our results.
The differences between \cite{Andersson} and \cite{Benhar}, attributed to the 
different sets of EOSs used in the two studies, are smaller than
the spread between the individual EOSs we considered.

For the damping times, also shown in \Fref{fig:fit1}, 
we fitted \Eref{eq:fit2} to our data, obtaining values
$a_2=0.084 \pm 0.012$, $b_2=-0.260$, where the error given for $a_2$ is the largest 
residual. The differences between our result and the fits from
\cite{Andersson} and \cite{Benhar} are smaller than the spread between the 
different EOSs, and thus compatible. However, as visible in \Fref{fig:fit1},
a linear relation 
does not describe the data well in the larger compactness range we consider.
Although this could be compensated by fitting a more appropriate curve, there 
is also a large spread between the different EOS. The spread and the deviation
from linearity strongly limits the usefulness of \Eref{eq:fit2} for 
constraining observational data.

\begin{table}
\begin{ruledtabular}
\caption{Parameters $a_1$, $b_1$ of the linear fit given by 
\Eref{eq:fit1} applied to each EOS individually, and to all
models combined. 
$\Delta \omega$ denotes the largest residual. 
For comparison, we also include the fits from \cite{Andersson} (AK Fit)
and \cite{Benhar} (BFG Fit).}
\begin{tabular}{lccc}
Model&
$a_1\, [\kilo\hertz]$&
$b_1\, [\kilo\hertz \usk\kilo\meter]$&
$\Delta \omega\, [\kilo\hertz]$
\\\hline 
AkmalPR&
$0.912$&
$29.05$&
$0.011$
\\
BalbN1H1&
$0.116$&
$52.08$&
$0.043$
\\
BBB2&
$0.705$&
$35.33$&
$0.013$
\\
BPAL12&
$0.355$&
$48.65$&
$0.001$
\\
FPS&
$0.634$&
$37.78$&
$0.005$
\\
GlendNH3&
$0.295$&
$45.79$&
$0.022$
\\
SHT&
$0.542$&
$32.98$&
$0.018$
\\
LS220&
$0.419$&
$40.65$&
$0.008$
\\
SLy4&
$0.713$&
$34.13$&
$0.008$
\\
Combined&
$0.332$&
$44.04$&
$0.275$
\\
BFG Fit&
$0.790$&
$33.00$&
---
\\
AK Fit&
$0.780$&
$35.96$&
---
\end{tabular}
\end{ruledtabular}
\label{tab:fit1_ind_eos}
\end{table}

\begin{figure*}[!htb]
\begin{center}
\includegraphics[width=\textwidth]{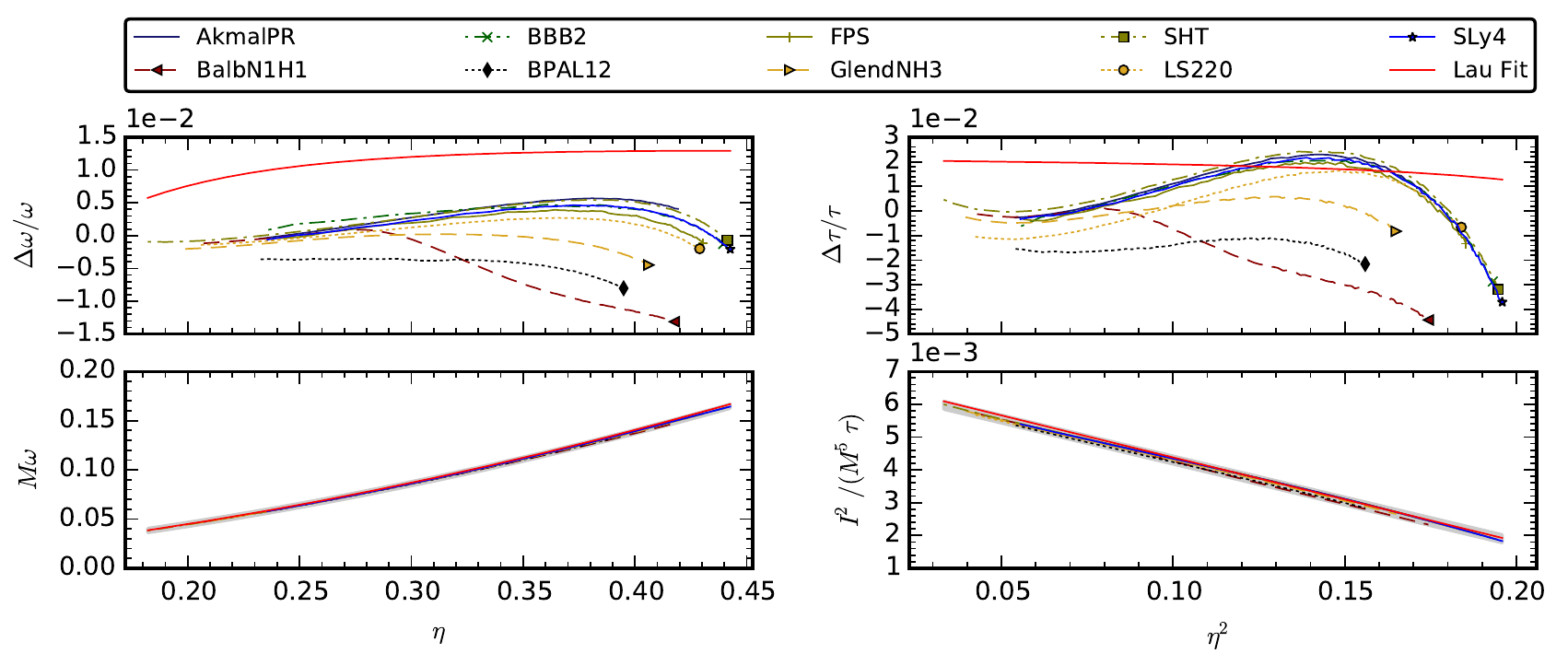}
\end{center}
\caption{Universal relations of the $f$ mode in terms of the effective compactness
and mass. The left panels show the relation \Eref{eq:fit3} for the frequency, 
the right panels the relation \Eref{eq:fit4} for the damping time.
The lower panes show the exact results, where the shaded area depicts the 
confidence bands containing all models (see main text).
The upper panels show the deviations from our fit. For comparison, we also plot the 
fits provided in Lau et al \cite{Lau}.
}
\label{fig:fit2}
\end{figure*}

Next, we tested the universal relations proposed in \cite{Lau}, which
express the $f$-mode properties in terms of the  effective compactness $\eta$.
In particular, we fit the expression 
\begin{align}
M\omega &= a_3 + b_3\eta + c_3\eta^2
\label{eq:fit3}
\end{align}
for the frequency to our results for all EOSs.
We find values 
$a_3 =-0.00383 \pm 0.0022$, $b_3 = 0.1276$, and $c_3 = 0.5718$.
The error provided for $a_3$ defines a confidence band which contains 
the results for all our models as well as the fit 
provided\footnote{Note there is a typo in Eq.~(6) from \cite{Lau}, where the 
last coefficient should read 0.575 instead of 0.0575 in order to be consistent 
with their Figure 1.} 
in \cite{Lau}. This takes into account that some of the EOSs used in 
\cite{Lau} are not part of our EOSs sample.
The fit results are shown in \Fref{fig:fit2}. As one can see, 
the universal relation  proposed in \cite{Lau} describes the results 
for our set of EOSs and models still very well. 

In order to parametrize the damping time, we fit the expression
\begin{align}
\frac{I^2}{M^5\tau} &= a_4 + b_4\eta^2 \,,
\label{eq:fit4}
\end{align}
proposed in \cite{Lau}, to our models. We obtain values
$a_4 = 0.00680 \pm 0.00013$ and $b_4 = -0.0250$.
The error given for $a_4$ again defines a band containing all our models as well as the 
fit provided in \cite{Lau}. 
The results are plotted in \Fref{fig:fit2}, showing that the 
universal relation for the damping time holds to good accuracy for our 
models.

\begin{figure}
\begin{center}
\includegraphics[width=\columnwidth]{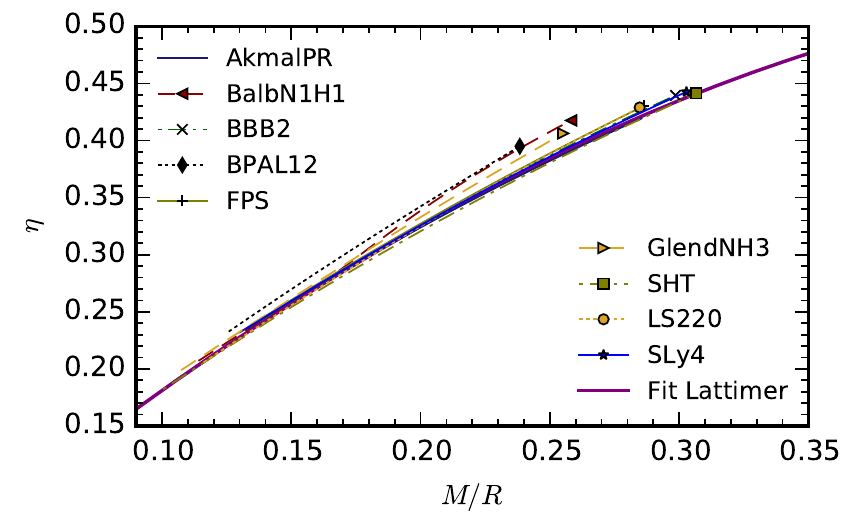}
\end{center}
\caption{Effective compactness $\eta$ versus compactness for 
all EOSs. In addition, we show the fit given in \cite{Lattimerschutz05}.
}
\label{fig:eta_comp}
\end{figure}

We note that the relation between compactness
and effective compactness is also weakly EOS dependent, as pointed 
out in \cite{Bejger, Lattimerschutz05}, where the normalized 
moment of inertia $\tilde{I} = I /(MR^2)$ is given as a function
of compactness $M/R$. This can easily be rearranged
in terms of effective compactness $\eta$ and compactness $M/R$.
For the EOSs considered here,
the resulting relation is shown in \Fref{fig:eta_comp}. 
Compared to the empirical fit proposed by \cite{Lattimerschutz05}, 
our models have a slightly larger effective compactness on average.

\begin{figure}
\begin{center}
\includegraphics[width=\columnwidth]{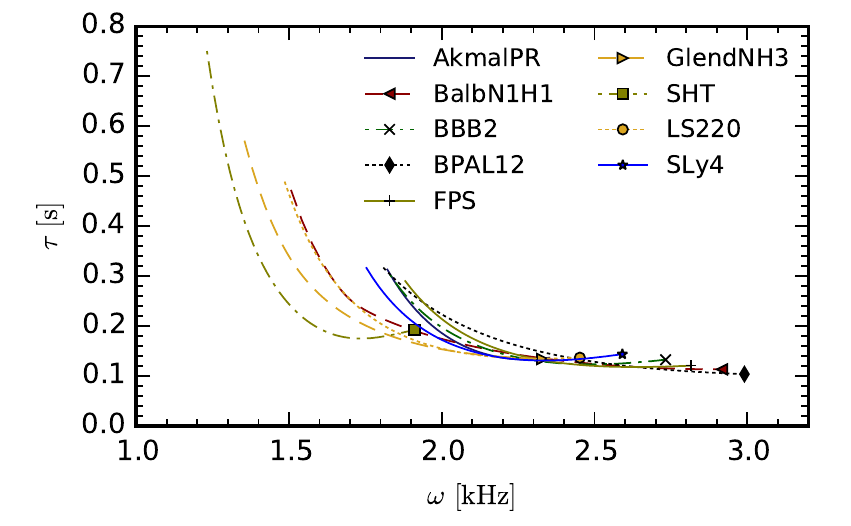}
\end{center}
\caption{Damping time due to GW emission versus oscillation frequency
for all EOSs. The symbols mark the maximum mass models, which also 
exhibit the largest frequency.}
\label{fig:omega_tau}
\end{figure}

In \cite{Lau}, it was proposed
to use the relations given by \Eqnsref{eq:fit3}{eq:fit4} to estimate 
mass and effective compactness from measurements of $\omega$ 
and $\tau$, and further to compute the radius using the aforementioned 
universal relation between $\eta$ and compactness.
The resulting mass and radius could in principle be used to constrain 
the EOS, although the remaining EOS dependency of the ``universal'' 
relations used for the estimate could be prohibitive.
To decide whether a simultaneous observation of frequency 
and damping time can really restrict the EOS, it is easier to simply 
plot the damping time as a function of frequency for all EOSs, as
shown in \Fref{fig:omega_tau}. As one can see, there is a considerable
overlap between the curves for different EOSs. 
Nevertheless, the maximum oscillation frequency, which is in general 
reached for the maximum mass model, depends strongly on the EOS. 
A measurement of a high frequency could thus rule out many EOSs,
even without knowledge of the damping time. The frequencies
of the maximum mass models can be found in the Appendix.

We also observe that the minimum damping time 
depends on the EOS. The measurement of a short damping timescale would
in principle constrain the EOS, even without knowledge of the frequency.  
In practice however, the damping is probably not caused by GW emission alone,
in particular at high amplitudes. Thus the observed damping time would only 
be a lower limit for the radiation reaction time, and hence not constrain the 
EOS.

\begin{figure}[!htb]
\begin{center}
\includegraphics[width=\columnwidth]{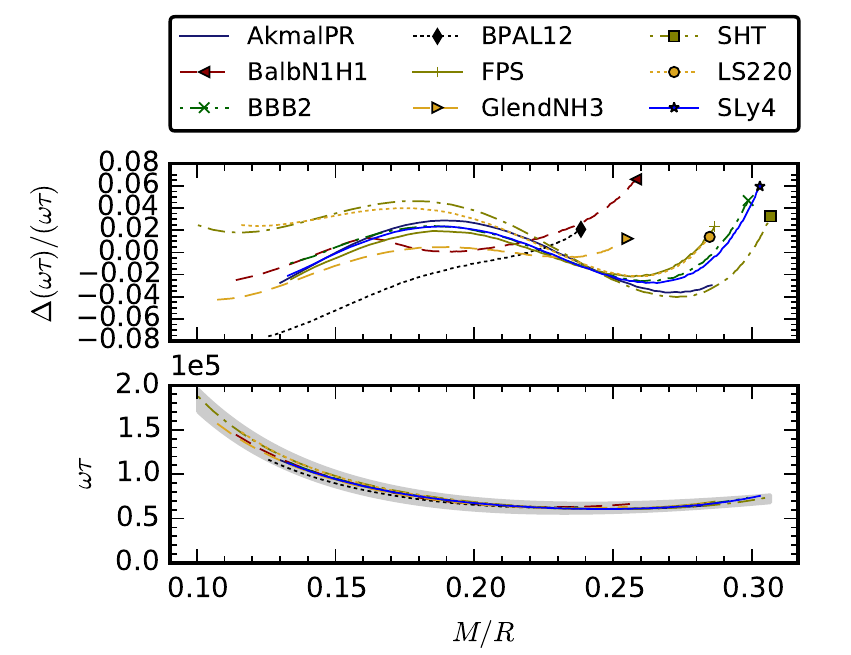}
\end{center}
\caption{Universal relation for the dimensionless quantity $\omega \tau$  
in terms of compactness. 
The lower panel shows the exact values for all EOSs together with 
the confidence band around the fitted value (see main text).
The upper panel shows the deviation from the fitted relation.}
\label{fig:fit3}
\end{figure}

Searching for a better relation for the damping time in terms of compactness, 
we turned to the 
dimensionless quantity $\omega\tau$. Multiplying by the frequency is
not a serious drawback since for any observation long enough to determine
the damping time, the frequency will also be known to good accuracy.
The choice is motivated by the fact that for a subset of EOS given 
by polytropic EOSs with fixed polytropic index, but arbitrary polytropic 
constant, all relations expressed in terms of quantities which are 
dimensionless in geometric units are automatically universal.
Note the fits given by \Eqnsref{eq:fit3}{eq:fit4} 
are already in terms of dimensionless quantities.
Although realistic EOS do not exhibit the underlying scaling invariance
of polytropes, we found that our data is well represented by the relation
\begin{align}\label{eq:fit5}
\omega \tau \frac{M}{R} 
&= a_5 + b_5 \frac{M}{R} + c_5 \left(\frac{M}{R}\right)^2
\end{align}
with $a_5= (3.69  \pm 0.129) \cdot 10^4$,
$b_5=-2.50 \cdot 10^5$, 
$c_5= 6.59 \cdot 10^5$. 
The error given for $a_5$ denotes 
the largest residual.
The fit is shown in \Fref{fig:fit3}. 
Although $\omega\tau$ is well constrained by the compactness,
the relation can only be inverted for low compactness.
It is therefore of limited use for constraining the compactness 
from a measurement of $\omega\tau$.

\section{Comparison with different methods}
\label{sec:compare}

In this section we study the accuracy of the Cowling approximation
with respect to $f$ modes and compare our results to three-dimensional nonlinear numerical 
simulations.

\subsection{Cowling approximation}

We now compare the frequencies obtained in full GR, 
presented in  \Sref{sec:results}, to those obtained using 
the Cowling approximation, which neglects the metric perturbations.
Keeping the gravitational field fixed leads to significant 
changes in the mode frequencies, see \cite{Yoshida, Dimmelmeier06}.
Further, there is no emission of gravitational waves and one obtains 
normal modes of oscillation, instead of decaying quasinormal modes.

In order to obtain the frequencies in the Cowling approximation,
we use a semi-analytic code case based on the linearized 
equations given in \cite{Kastaun08}, assuming harmonic time dependence 
and specializing to spherical symmetry. 
The resulting singular boundary value problem for the eigenfunctions
is solved using the method of Frobenius to compute the solution at 
the boundaries (origin and surface) in conjunction with a shooting 
method to determine the mode frequency. 

\begin{figure}[!htb]
\begin{center}
\includegraphics[width=0.95\linewidth]{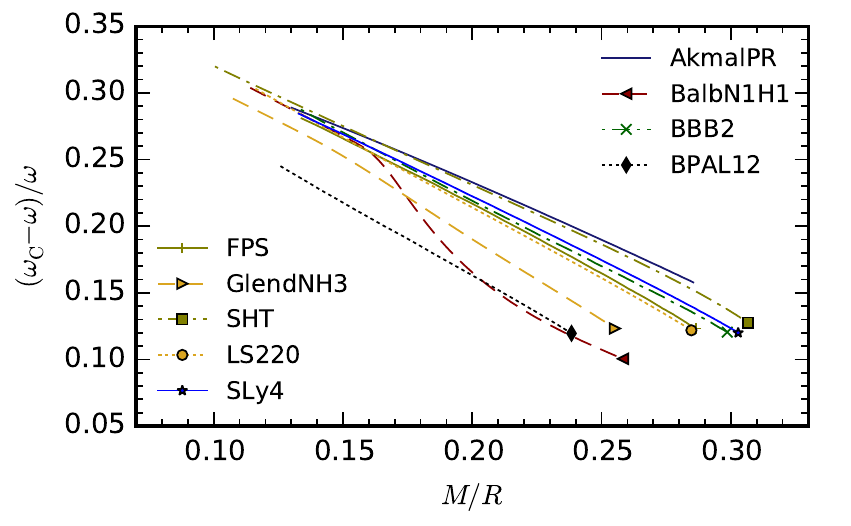}
\end{center}
\caption{The relative difference between $f$-mode frequencies obtained
in full GR and the Cowling approximation.}
\label{fig:diff_cow}
\end{figure}

The relative difference for all the EOS considered here is displayed in  
\Fref{fig:diff_cow}. 
Our comparison matches the conclusions presented in \cite{Yoshida}: 
the difference ranges from ${\approx}10$ -- $30\%$ 
and decreases with increasing stellar compactness. A possible explanation for this trend was proposed by \cite{Shin}, noting that increasing compactness can make the role of metric perturbations less relevant for the $f$-mode eigenfunction, given that the eigenfunction is peaked near the surface.

\subsection{Numerical relativity simulations}

As an additional cross check, 
we compare our linear results (both GR and Cowling) to those obtained 
with a nonlinear, three-dimensional numerical evolution code.
The nonlinear results are more expensive and, due to limited
resolution, less accurate, but they are completely independent 
both regarding the analytic formalism and the numerical implementation.
Note that the numerical damping typical for such codes prevents
the computation of the physical damping timescale associated 
with GW radiation, which is why we only compare the frequencies.

The nonlinear hydrodynamic evolution code is described in 
\cite{Galeazzi2013}. It is based on a conservative formulation of 
the hydrodynamic evolution equations, which are solved using finite 
volume, high-resolution-shock-capturing methods. In particular, it 
employs piecewise parabolic reconstruction in conjunction with the 
HLLE approximate Riemann solver to compute fluxes. 
The spacetime is evolved by means of the publicly available McLachlan 
code \cite{Brown2009}, using the CCZ4 formulation of the metric 
evolution equations \cite{Alic2012, Alic2013}.
Note that we do not consider composition effects here, i.e., we 
ignore the electron fraction altogether and assume that the barotropic EOS 
which describes the background model is also valid during the 
evolution. The same assumption is implicitly made for the linear 
perturbation code.

The simulations made use of nested-box fixed mesh refinement with 
6 refinement levels centered around the star. In this setup, the finest 
level fully contains the star, and the grid spacing is $222 \usk\meter$.
We employ radiative boundary conditions for the metric at the outer 
boundary, which is located at $708 \usk\kilo\meter$.

\begin{figure}
\begin{center}
\includegraphics[width=0.95\columnwidth]{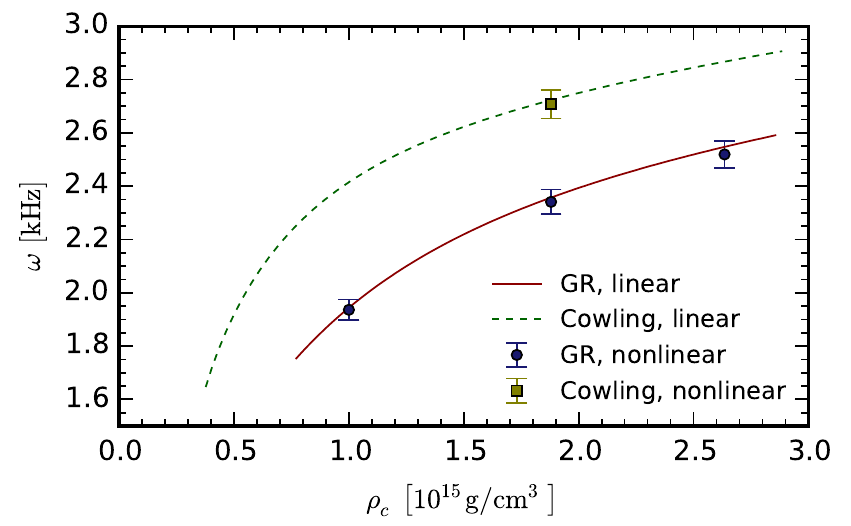}
\end{center}
\caption{The $f$-mode frequency $\omega$ for the SLy4 EOS, calculated 
with the linear codes in full GR and within the Cowling approximation, 
in comparison to results obtained with the three-dimensional nonlinear 
evolution code.}
\label{fig:freq_numrel}
\end{figure}

Using this setup, we compared frequencies for three models with 
different central densities, all obeying the SLy EOS (chosen as a representative case). 
The results are shown in \Fref{fig:freq_numrel}.
We estimate the numerical errors (due to the numerical evolution and 
the resolution of the Fourier analysis used to extract the frequency)
to be around 2\%. The maximum difference between the frequencies obtained 
with the two codes is 1.1\%.  The results thus agree within the expected
accuracy.
We also computed the frequency for one model in the Cowling approximation.
As shown in \Fref{fig:freq_numrel}, the result matches the one using
the linear code within 0.6\%.

\section{Conclusions and final remarks}
\label{sec:conclusions}

In this work, we provided accurate results for the $f$ modes of neutron stars 
described by the 9 EOSs listed in \Tref{tableEOS}. 
Our results were obtained by solving the linear perturbation equations that describe the nonradial oscillations of relativistic neutron stars \cite{Lindblom1983,Lindblom1985}. 

With these results, we were able to update the empirical fits proposed in the literature that describe the general behavior of the $f$-mode frequency $\omega$ and damping time $\tau$ as functions of the star's average density and compactness \cite{Andersson, Benhar}, based on the behavior of Newtonian stars and on the quadrupole formula, respectively. 

Our results cover a comprehensive range of masses and contain more modern EOSs,
allowing us to test proposed universal relations more stringently. 
In particular, we clearly showed that $\tau$ does not follow the linear fit 
proposed by \cite{Andersson}. 
We did however find a new universal relation for the quantity $\omega\tau$ as a 
function of the compactness, which describes our results more accurately.
It can be used to restrict $\omega\tau$ from the compactness, but can only 
be inverted for low compactness.
The more recent proposal in \cite{Lau} for a universal relations describing 
the $f$ modes in terms of the effective compactness still holds to very good 
accuracy for our set of EOS.

In our comparison with the results of a linear Cowling code, we confirmed previous results from \cite{Yoshida} and extended them to more EOSs. The difference between the Cowling and the GR results ranges from ${\approx}10$ -- $30\%$ and decreases with increasing compactness.

Finally we compared our results with a three-dimensional nonlinear evolution code for a few stellar models employing the SLy4 EOS. We showed that the results agree well within the estimated error bars for the numerical results, thus validating the implementation of the different approaches presented here.

\begin{acknowledgments}
This work was supported by CAPES and the Max Planck Society. Some of the computations 
were carried out on the Datura cluster of the AEI. We thank F. Galeazzi for preparing 
the LS220 and SHT EOS tables. C. Chirenti wishes to thank L. Rezzolla  and 
S. Yoshida for useful discussions.
\end{acknowledgments}

\bigskip

\appendix*
\section{Numerical results for each EOS}
\label{A1}

In this appendix we present first numerical results for the $f$-mode frequency $\omega$ and damping time $\tau$ for a few representative stars generated with each one of the 9 EOS considered in this paper (see \Tref{tableEOS}). In addition, we provide
mass, radius, and effective compactness $\eta$.
These results are given in tables \ref{tab:modelsAkmalPR}--\ref{tab:modelsSLy4}.

\begin{table}
\begin{ruledtabular}
\caption{Properties of selected models for the AkmalPR EOS.
$\rho_c$ is the central energy density,
$M$ is the gravitational mass, $R$ the circumferential radius,
$\eta$ the effective compactness (dimensionless), $\omega$ is the 
frequency of the $m=0, l=2$ $f$ mode, $\tau$ the corresponding
damping timescale due to gravitational radiation.
The first line refers to the most massive model.}
\label{tab:modelsAkmalPR}
\begin{tabular}{cccccc}
$\rho_c\, [10^{15}\gram\per\centi\meter\cubed]$&
$M\, [M_\odot]$&
$R\, [\kilo\meter]$&
$\eta$&
$\omega\, [\kilo\hertz]$&
$\tau\, [\second]$
\\\hline 
$1.771$&
$2.071$&
$10.716$&
$0.419$&
$2.352$&
$0.137$
\\
$1.315$&
$1.796$&
$11.141$&
$0.369$&
$2.189$&
$0.138$
\\
$1.097$&
$1.536$&
$11.296$&
$0.325$&
$2.069$&
$0.161$
\\
$0.939$&
$1.276$&
$11.358$&
$0.280$&
$1.953$&
$0.209$
\\
$0.800$&
$1.003$&
$11.382$&
$0.232$&
$1.822$&
$0.313$
\end{tabular}
\end{ruledtabular}
\end{table}

\begin{table}
\begin{ruledtabular}
\caption{Like table \ref{tab:modelsAkmalPR}, but for the BalbN1H1 EOS}
\label{tab:modelsBalbN1H1}
\begin{tabular}{cccccc}
$\rho_c\, [10^{15}\gram\per\centi\meter\cubed]$&
$M\, [M_\odot]$&
$R\, [\kilo\meter]$&
$\eta$&
$\omega\, [\kilo\hertz]$&
$\tau\, [\second]$
\\\hline 
$3.791$&
$1.631$&
$9.323$&
$0.418$&
$2.917$&
$0.113$
\\
$1.311$&
$1.474$&
$12.448$&
$0.297$&
$1.850$&
$0.204$
\\
$0.731$&
$1.326$&
$12.943$&
$0.260$&
$1.662$&
$0.278$
\\
$0.634$&
$1.151$&
$13.017$&
$0.231$&
$1.576$&
$0.365$
\\
$0.570$&
$1.009$&
$13.058$&
$0.207$&
$1.507$&
$0.473$
\end{tabular}
\end{ruledtabular}
\end{table}

\begin{table}
\begin{ruledtabular}
\caption{Like table \ref{tab:modelsAkmalPR}, but for the BBB2 EOS}
\label{tab:modelsBBB2}
\begin{tabular}{cccccc}
$\rho_c\, [10^{15}\gram\per\centi\meter\cubed]$&
$M\, [M_\odot]$&
$R\, [\kilo\meter]$&
$\eta$&
$\omega\, [\kilo\hertz]$&
$\tau\, [\second]$
\\\hline 
$3.189$&
$1.921$&
$9.500$&
$0.439$&
$2.733$&
$0.133$
\\
$1.619$&
$1.696$&
$10.686$&
$0.369$&
$2.313$&
$0.130$
\\
$1.245$&
$1.457$&
$11.066$&
$0.319$&
$2.114$&
$0.161$
\\
$1.011$&
$1.231$&
$11.260$&
$0.275$&
$1.958$&
$0.216$
\\
$0.847$&
$1.028$&
$11.350$&
$0.236$&
$1.830$&
$0.303$
\end{tabular}
\end{ruledtabular}
\end{table}

\begin{table}
\begin{ruledtabular}
\caption{Like table \ref{tab:modelsAkmalPR}, but for the BPAL12 EOS}
\label{tab:modelsBPAL12}
\begin{tabular}{cccccc}
$\rho_c\, [10^{15}\gram\per\centi\meter\cubed]$&
$M\, [M_\odot]$&
$R\, [\kilo\meter]$&
$\eta$&
$\omega\, [\kilo\hertz]$&
$\tau\, [\second]$
\\\hline 
$3.981$&
$1.455$&
$9.013$&
$0.395$&
$2.991$&
$0.104$
\\
$1.954$&
$1.339$&
$10.515$&
$0.325$&
$2.362$&
$0.144$
\\
$1.477$&
$1.223$&
$11.102$&
$0.289$&
$2.122$&
$0.187$
\\
$1.209$&
$1.120$&
$11.504$&
$0.260$&
$1.958$&
$0.240$
\\
$1.000$&
$1.011$&
$11.866$&
$0.233$&
$1.809$&
$0.317$
\end{tabular}
\end{ruledtabular}
\end{table}

\begin{table}
\begin{ruledtabular}
\caption{Like table \ref{tab:modelsAkmalPR}, but for the FPS EOS}
\label{tab:modelsFPS}
\begin{tabular}{cccccc}
$\rho_c\, [10^{15}\gram\per\centi\meter\cubed]$&
$M\, [M_\odot]$&
$R\, [\kilo\meter]$&
$\eta$&
$\omega\, [\kilo\hertz]$&
$\tau\, [\second]$
\\\hline 
$3.384$&
$1.800$&
$9.280$&
$0.431$&
$2.816$&
$0.121$
\\
$1.672$&
$1.598$&
$10.515$&
$0.358$&
$2.335$&
$0.131$
\\
$1.299$&
$1.402$&
$10.845$&
$0.315$&
$2.155$&
$0.161$
\\
$1.076$&
$1.212$&
$11.023$&
$0.277$&
$2.016$&
$0.207$
\\
$0.902$&
$1.011$&
$11.139$&
$0.238$&
$1.881$&
$0.290$
\end{tabular}
\end{ruledtabular}
\end{table}

\begin{table}
\begin{ruledtabular}
\caption{Like table \ref{tab:modelsAkmalPR}, but for the GlendNH3 EOS}
\label{tab:modelsGlendNH3}
\begin{tabular}{cccccc}
$\rho_c\, [10^{15}\gram\per\centi\meter\cubed]$&
$M\, [M_\odot]$&
$R\, [\kilo\meter]$&
$\eta$&
$\omega\, [\kilo\hertz]$&
$\tau\, [\second]$
\\\hline 
$2.375$&
$1.966$&
$11.378$&
$0.406$&
$2.329$&
$0.134$
\\
$1.046$&
$1.729$&
$13.357$&
$0.321$&
$1.790$&
$0.190$
\\
$0.743$&
$1.492$&
$13.993$&
$0.272$&
$1.587$&
$0.271$
\\
$0.553$&
$1.232$&
$14.310$&
$0.228$&
$1.443$&
$0.409$
\\
$0.477$&
$1.051$&
$14.457$&
$0.199$&
$1.354$&
$0.571$
\end{tabular}
\end{ruledtabular}
\end{table}

\begin{table}
\begin{ruledtabular}
\caption{Like table \ref{tab:modelsAkmalPR}, but for the LS220 EOS}
\label{tab:modelsLS220}
\begin{tabular}{cccccc}
$\rho_c\, [10^{15}\gram\per\centi\meter\cubed]$&
$M\, [M_\odot]$&
$R\, [\kilo\meter]$&
$\eta$&
$\omega\, [\kilo\hertz]$&
$\tau\, [\second]$
\\\hline 
$2.574$&
$2.053$&
$10.646$&
$0.429$&
$2.452$&
$0.137$
\\
$1.191$&
$1.788$&
$12.223$&
$0.346$&
$1.969$&
$0.160$
\\
$0.891$&
$1.525$&
$12.636$&
$0.295$&
$1.777$&
$0.214$
\\
$0.710$&
$1.273$&
$12.824$&
$0.251$&
$1.628$&
$0.306$
\\
$0.570$&
$1.014$&
$12.891$&
$0.206$&
$1.486$&
$0.489$
\end{tabular}
\end{ruledtabular}
\end{table}

\begin{table}
\begin{ruledtabular}
\caption{Like table \ref{tab:modelsAkmalPR}, but for the SHT EOS}
\label{tab:modelsSHT}
\begin{tabular}{cccccc}
$\rho_c\, [10^{15}\gram\per\centi\meter\cubed]$&
$M\, [M_\odot]$&
$R\, [\kilo\meter]$&
$\eta$&
$\omega\, [\kilo\hertz]$&
$\tau\, [\second]$
\\\hline 
$1.554$&
$2.769$&
$13.334$&
$0.441$&
$1.911$&
$0.193$
\\
$0.738$&
$2.316$&
$14.719$&
$0.360$&
$1.630$&
$0.186$
\\
$0.584$&
$1.896$&
$14.920$&
$0.305$&
$1.507$&
$0.240$
\\
$0.466$&
$1.427$&
$14.917$&
$0.242$&
$1.369$&
$0.385$
\\
$0.371$&
$1.006$&
$14.788$&
$0.182$&
$1.232$&
$0.750$
\end{tabular}
\end{ruledtabular}
\end{table}

\begin{table}
\begin{ruledtabular}
\caption{Like table \ref{tab:modelsAkmalPR}, but for the SLy4 EOS}
\label{tab:modelsSLy4}
\begin{tabular}{cccccc}
$\rho_c\, [10^{15}\gram\per\centi\meter\cubed]$&
$M\, [M_\odot]$&
$R\, [\kilo\meter]$&
$\eta$&
$\omega\, [\kilo\hertz]$&
$\tau\, [\second]$
\\\hline 
$2.856$&
$2.049$&
$9.992$&
$0.443$&
$2.591$&
$0.144$
\\
$1.417$&
$1.793$&
$11.284$&
$0.368$&
$2.183$&
$0.138$
\\
$1.084$&
$1.517$&
$11.635$&
$0.316$&
$2.001$&
$0.172$
\\
$0.896$&
$1.273$&
$11.791$&
$0.272$&
$1.865$&
$0.230$
\\
$0.771$&
$1.068$&
$11.866$&
$0.236$&
$1.754$&
$0.316$
\end{tabular}
\end{ruledtabular}
\end{table}

\end{document}